\documentclass[twocolumn,pra]{revtex4-1}
\usepackage{braket}
\usepackage{physics}
\usepackage{xfrac}

\begin{document}

\title{Size invariant twisted optical modes for efficient generation of higher dimensional quantum states}

\author{Ali Anwar$^{1,2,*}$, Shashi Prabhakar$^{3}$, R. P. Singh$^{1}$}
\affiliation{$^{1}$Quantum Science and Technology Laboratory, Physical Research Laboratory, Ahmedabad, Gujarat, India}
\affiliation{$^{2}$Centre for Quantum Technologies, National University of Singapore, Singapore}
\affiliation{$^{3}$Photonics Laboratory, Physics Unit, Tampere University, Tampere, FI-33720, Finland}
\affiliation{$^{*}$Email: alianwarma@gmail.com}

\date{\today}

\begin{abstract}
Optical vortex beams are profiled as helical wavefronts with a phase singularity carrying an orbital angular momentum (OAM) associated with their spatial distribution. The transverse intensity distribution of a conventional optical vortex has a strong dependence on the carried topological charge. However, perfect optical vortex (POV) beams have their transverse intensity distribution independent of their charge. Such `size-invariant' POV beams have found exciting applications in optical manipulation, imaging and communication. In this article, we investigate the use of POV modes in the efficient generation of high dimensional quantum states of light. We generate heralded single photons carrying OAM using spontaneous parametric down-conversion (SPDC) of POV beams. We show that the heralding efficiency of the SPDC single photons generated with POV pump is greater than that with normal optical vortex beams. The dimensionality of the two-photon OAM states is increased with POV modes in the pump and projective measurements using Bessel-Gaussian vortex modes that give POV, instead of the Laguerre-Gaussian modes.
\end{abstract}

\maketitle

\section{Introduction}
Light carrying OAM, which corresponds to beams with a helical twist of phase, has gained interest due to its applications in various fields of optics. Generally termed as `Optical Vortex' in classical optics, light having phase singularity has been used in the generation of bright optical solitons \cite{swartzlander1, tikhonenko}, study of optical chronograph \cite{swartzlander2, berkhout} and trapping of particles \cite{gahagan}. The transverse intensity distribution of an optical vortex mode is radially symmetric in nature that originates from the azimuthal dependence of the phase. Among such classes of phase singular modes, Laguerre-Gaussian (LG) modes are most commonly used in practical applications due to their profile stability upon free space propagation. Since the size of a normal optical vortex strongly depends on its topological charge \cite{reddydivergence}, they have their limitations in applications involving transmission of OAM modes through optical fibers \cite{li, gregg} in communication \cite{willner}. Due to this, projective measurements based on `phase-flattening' technique become difficult for higher-order orbital angular momentum (OAM) modes \cite{qassim}. Bessel-Gaussian (BG) modes are another class of structured light modes carrying OAMs, which come from the non-diffracting solution of paraxial wave equation \cite{gori1987bessel, mcgloin2005bessel}. The transverse field distribution of BG modes is described by a Bessel function \cite{korenev2002bessel}. They exhibit non-diffracting nature \cite{durnin1987diffraction, cruz2012observation} as well as self-healing property \cite{litvin2009conical, mclaren2014self} when propagated in space. In most of the practical cases, Bessel-Gaussian modes are generated using axicons \cite{arlt2000generation} and spatial light modulators (SLM) \cite{leach2006generation}. Fourier transformation of BG modes gives a class of size invariant modes known as Perfect optical vortex (POV) \cite{vaity2015}. They are a new class of ring-shaped beams carrying OAM. The size of a POV modes does not change with respect to the change in OAM values.

The concept of OAM has been well established in quantum information too. Access to higher dimensions in Hilbert space makes OAM suitable for encoding higher amounts of information per photon. SPDC is the most common workhorse for the generation of single photons carrying OAM \cite{boydbook, mair}. The paired photons generated in SPDC are inherently correlated in their OAM values \cite{walborn2004entanglement}. It has been experimentally verified that the amplitude, as well as the helical phase of an optical vortex pump, gets transferred to SPDC photons \cite{vicuna, anwar2018direct}. Based on the OAM selection rule in SPDC process, single photons carrying OAM are generated in `heralding’ configuration by pumping an optical vortex beam into a nonlinear crystal and projecting one photon from the generated pair to `zero OAM' (Gaussian) mode \cite{lal2020photon}. Single-photons carrying OAM find potential applications in quantum information \cite{erhard}, quantum gates \cite{babazadeh}, quantum memories \cite{nicolas} etc. 

Apart from the Gaussian modes, OAM correlations among the SPDC photon-pairs are explored using structured pump modes \cite{romero}. Such systems can be used to generate versatile high-dimensional quantum states by careful shaping of pump beam having additional spiral modes using different pump engineering techniques \cite{liu2018coherent, kovlakov2018quantum, anwar2020selective}. Also, it was shown that the use of BG modes in photon-pair detection using OAM projection improved the dimensionality of the OAM state generated. However, the potential of the class of such modes in further enhancement of the heralded single-photon detection and dimensionality of the quantum state is yet unexplored. In this article, we discuss the use of POV modes as the pump in a parametric down-conversion process to generate heralded single photons carrying OAM. By exploiting the OAM independence on the size of POV modes, we show that the heralding efficiency of the single photons with POV pump is higher than that with a NOV pump. We also show that a configuration with POV pump beams with generated SPDC photon-pairs projected to BG modes give improved fidelity than the configuration with an LG mode projection.

\section{Helical modes and their generations} \label{gen-non-diffr-size-inv-modes}
Depending on the transverse intensity distribution, there are three main classes of optical modes carrying azhimuthal phase. All those modes have phase singularity at the center where the intensity is zero. Here we introduce different helical modes and an experimental method to generate them.
\begin{figure}[h]
	\centering
    \includegraphics[width=0.46\textwidth]{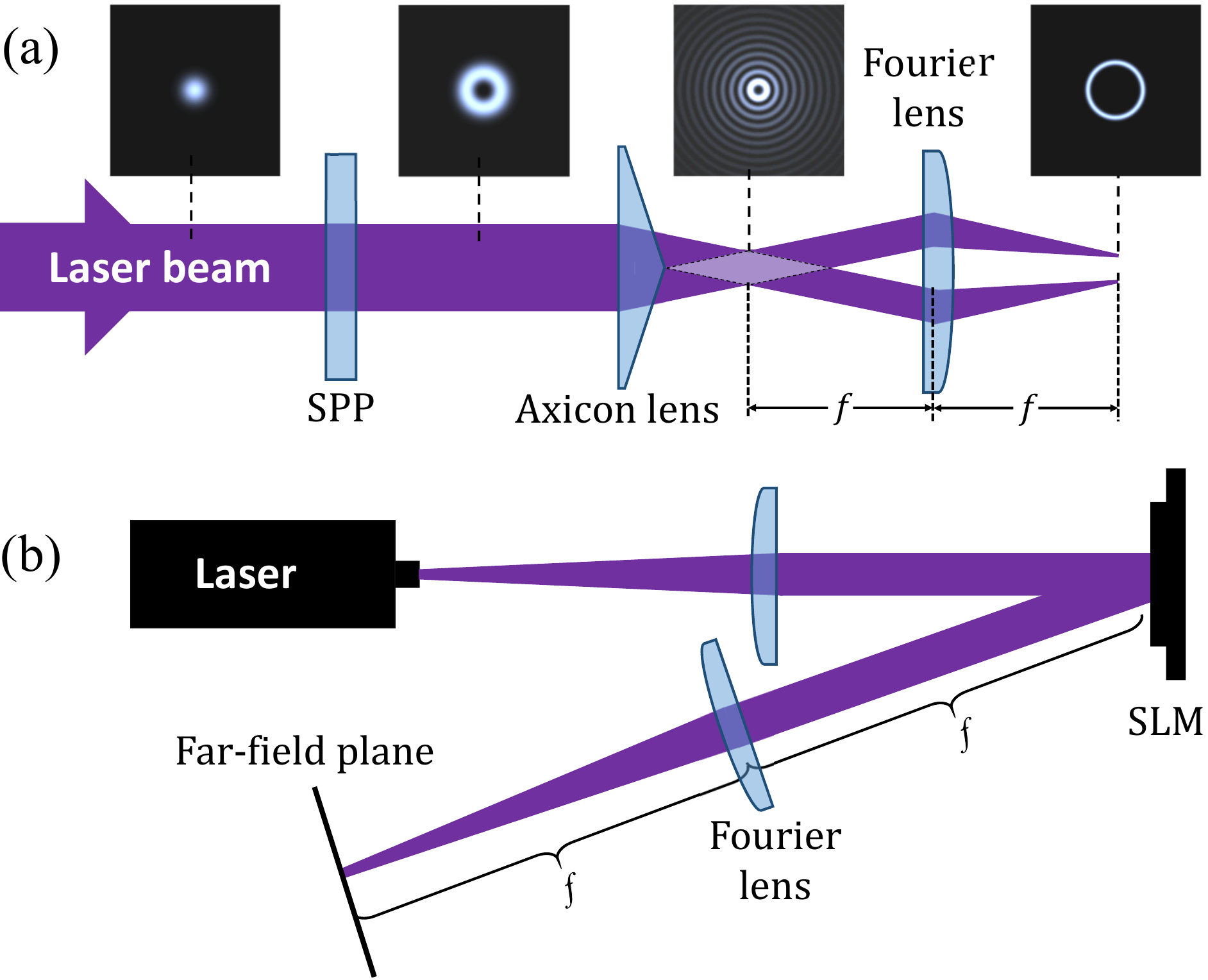}
	\caption{Experimental schemes to generate NOV, then BGV, and then POV: (a) Conversion of a Gaussian beam to helical modes Using spiral phase plate, axicon lens and a spherical lens. (b) Generation of helical modes by diffraction of fork grating hologram imprinted on a SLM.} \label{pump_prep_POV}
\end{figure}

\subsection{Size-variant: Normal optical vortex (NOV)}
\begin{figure}[h]
  \centering
  \includegraphics[width=0.46\textwidth]{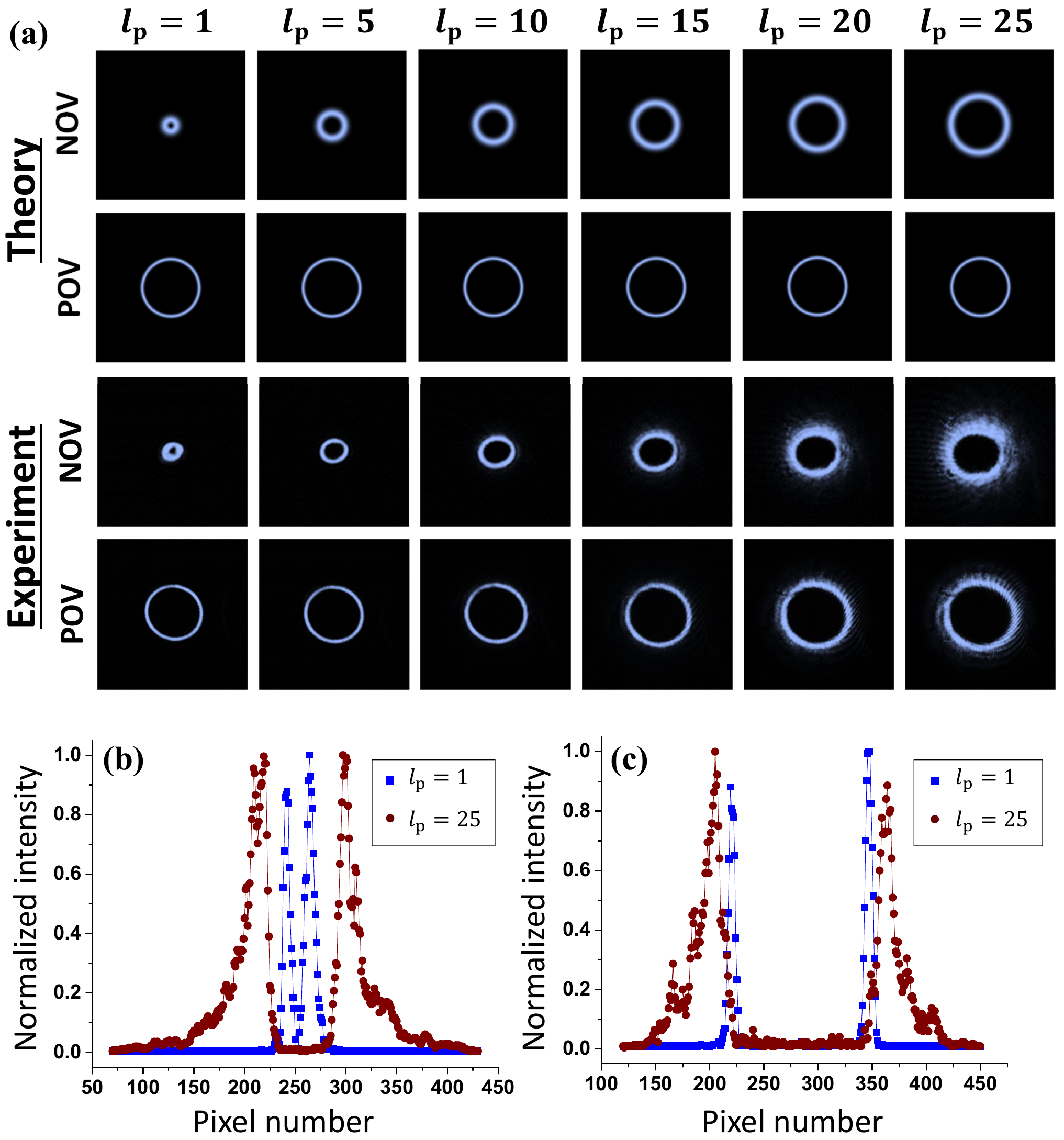}
  \caption{(a) Theoretical (top) and experimental (bottom) intensity distributions of NOV and POV beams of order varying from $\ell=1$ to $25$. (b) and (c) compares the intensity distribution
of a NOV and POV with $\ell=1$ and $\ell=25$.}\label{pumpmodes}
\end{figure}

A typical field distribution of a normal optical vortex (NOV) mode of topological charge $\ell$ in polar coordinates $(r,\theta)$, is given by
\begin{equation}
E_{\text{NOV}}^{\ell}(r,\theta)=\sqrt{\frac{2^{\abs{\ell}+1}}{\pi w^2|\ell|!}}\left(\frac{r}{w}\right)^{|\ell|}\exp\left(-\frac{r^2}{w^2}\right)\exp(i\ell\theta),
\label{NOV}
\end{equation}
where $w$ is the radius of the mode. As shown in Fig. \ref{pumpmodes}, the size of NOV modes increases with OAM due to the OAM-dependant radial term $r^\abs{\ell}$ in the above expression.

\subsection{Size-invariant: Perfect optical vortex (POV)}
A new class of optical vortex beam, termed as `perfect optical vortex', was introduced by Ostrovsky \textit{et. al.} that solves size effects of a normal vortex \cite{ostrovskygeneration}. Conventionally, POV of order $\ell$ are formed as a Fourier transform of Bessel-Gaussian vortex (BGV) of order $\ell$ \cite{vaity2015}. The normalized expression for a BGV mode of azhimuthal order $\ell$ is written as
\begin{equation}
E_{\rm{BG}}^{\ell}(r,\theta)=\sqrt{\frac{2e^{\sfrac{1}{4}}}{\pi w^2I_\ell(\sfrac{1}{4})}}J_\ell(k_rr)\exp\left(-\frac{r^2}{w^2}\right)\exp(i\ell\theta).
\label{BG}
\end{equation}
The Fourier transform of the above equation leads to the field of a typical perfect optical vortex of order $\ell$ is given as
\begin{equation}
E_{\rm{POV}}^{\ell}(r,\theta)=i^{\ell-1}\frac{w_g}{w_o}\exp(i\ell\theta)\exp\left(-\frac{(r^2+r_r^2)}{w_o^2}\right)I_\ell\left(\frac{2r_rr}{w_o^2}\right),
\label{POV-expression}
\end{equation}
where $w_g$ is the waist radius of the initial Gaussian beam, $w_o=2f/kw_g$ is half of the ring width and $r_r$ is radius of the ring. Here $f$ is the focal length of the Fourier lens and $k$ is the magnitude of the wave-vector of the light beam. 

\subsection{Experimental generation of NOV and POV modes}
Figure \ref{pump_prep_POV}(a) shows a simple method to generate POV beams using a spiral phase plate (SPP), an axicon, and a lens. A Gaussian beam passing through an SPP of topological order $\ell$ generates an NOV of OAM $\ell$. This beam then passing through the axicon becomes a BGV, and Fourier transformation of the BGV using lens generates a POV beam. Alternatively, these helical modes can also be generated from the diffraction of a laser beam using a fork-grating pattern imprinted onto SLM, as shown in Fig. \ref{pump_prep_POV}(b). The first diffracted order is imaged in the far-field plane will contain the desired helical mode, or NOV modes. A Gaussian light incident on a fork-grating hologram obtained through interference converts into BG modes in the immediate plane after SLM, and the corresponding POV mode is generated at the far-field plane.

In Fig \ref{pumpmodes}(a), the experimental intensity distribution of normal optical vortex pump is compared with the perfect optical vortex pump of different orders $\ell$=1, 5, 10, 15, 20 and 25. For a NOV pump, Eqn.1 shows an $\ell$ dependence on the radial terms. So, the size of the OAM modes increases with $\ell$. However, for a POV pump, radial terms are independent of $\ell$ (\eqref{BG}), and thus the size of modes remains the same for higher OAM values. The spatial profiles are shown in Fig \ref{pumpmodes}(b) and (c) for NOV and POV, respectively. From the intensity profile, one can observe that the spatial profile change considerably for a NOV, while remains unchanged for POVs.

\section{Parametric down conversion of helical modes} \label{PDC}
First, we will discuss the theory of SPDC with NOV and POV pump beams. A comparative study of down-converted photons with these pump modes is important to understand their two-photon modal spectra.

\subsection{Theory} \label{PDC:theory}
In the perturbative treatment of spontaneous parametric down-conversion process, interaction of pump ($p$), signal ($s$) and idler ($i$) modes in a medium (Non-linear $\chi^{(2)}$ crystal) is represented by an interaction Hamiltonian $\mathcal{H}_I$. The initial state is a vacuum state $\vert0\rangle_s\vert0\rangle_i$, therefore the output state of SPDC is approximated as
\begin{equation}
\vert\Phi\rangle\approx\left(1-\frac{i}{\hbar}\int_{0}^{\tau}\mathcal{H}_I(t)dt\right)\vert0\rangle_s\vert0\rangle_i.
\label{biphotonstate}
\end{equation}
The biphoton mode function of the generated twin photons in transverse momentum coordinates ($\mathbf{k}$) is obtained as
\begin{equation}
\Phi(\mathbf{k}^{\perp})=\langle\mathbf{k}^{\perp}_s\vert\langle-\mathbf{k}^{\perp}_i
\vert\Phi\rangle,
\label{biphotonmode}
\end{equation}
where $\mathbf{k}^{\perp}_s$ and $-\mathbf{k}^{\perp}_i$ represents the transverse position in the momentum coordinates for signal and idler respectively. On simplification, the biphoton mode 
function in transverse momentum coordinates is given by
\begin{equation}
\Phi(\mathbf{k}_s^{\perp},\mathbf{k}_i^{\perp},\Delta k)=E_p(\mathbf{k}_p^{\perp})L\text{sinc}\left(\dfrac{\Delta kL}{2}\right)\exp\left(i\dfrac{\Delta kL}{2}\right),
\label{modefna}
\end{equation}
where $E_p(\mathbf{k}_p^{\perp})$ represents the pump transverse amplitude distribution, $\mathbf{k}_p^{\perp} (= \mathbf{k}^{\perp}_s+\mathbf{k}^{\perp}_i)$ is the angular coordinates of the pump, $L$ is the thickness of the crystal and the exponential factor in the Eqn. \ref{modefna} is a global phase term. $\Delta k$ is the longitudinal phase mismatch given by
\begin{equation}
    \Delta k(\mathbf{k}_s^{\perp},\mathbf{k}_i^{\perp}) = k_p^z(\mathbf{k}_s^{\perp},\mathbf{k}_i^{\perp})-k_s^z(\mathbf{k}_s^{\perp})-k_i^z(\mathbf{k}_i^{\perp}).
    \label{phase_mismatch}
\end{equation}

Consider a Type-0/I SPDC using a $\chi^{(2)}$ crystal of thickness $L$. The magnitude of the wave-vector of the interacting fields under phase-matching condition are given by
\begin{align}
k_{p,s,i}^z(\mathbf{k}_{p,s,i}^{\perp})&=\sqrt{k_{p,s,i}^2-\vert \mathbf{k}_{p,s,i}^{\perp}\vert^2}.
\end{align}
Here $k_j$=$n_j\omega_j/c$ $(j=p,s,i)$ are the magnitudes of wave-vectors of the fields.  Here,  $n_j\equiv n_j(\omega_j)$ $(j=p,s,i)$ are the refractive indices and $\omega_j$ are the frequencies of the pump, signal and idler respectively. $c$ is the speed to light in vacuum. For near-collinear SPDC, the z-component of the wave-vector is approximated by $k-|\mathbf{k}^{\perp}|^2/2k$. So, \eqref{phase_mismatch} becomes
\begin{align}
    \Delta k(\mathbf{k}_s^{\perp},\mathbf{k}_i^{\perp}) = \frac{|\mathbf{k}_s^{\perp}-\mathbf{k}_i^{\perp}|^2}{2k_p}.
    \label{phase-mismatch-approx}
\end{align}
Also, an additional term of $-2\pi/\Lambda$ will be added to the phase-mismatch in \eqref{phase-mismatch-approx}, in case of a quasi phase-matched crystal, where $\Lambda$ is the grating period.

\subsection{Angular spectrum of SPDC} \label{PDC:experiment}
The generated signal and idler photons in SPDC propagate in space according to the phase-matching condition of SPDC. The locus of all points in space that satisfies phase-matching conditions comes out to be an annular ring distribution where the signal and idler photons in a pair are present. The angular spectrum of the down-converted signal photons for frequency $\omega_s$ is obtained by tracing the biphoton mode function overall idler photons \cite{yasser}
\begin{equation}
    R_s(\mathbf{k}^{\perp}_s)=\int 
    d\mathbf{k}_i^{\perp}\vert\Phi(\mathbf{k}_s^{\perp},\mathbf{k}_i^{\perp},\Delta k)\vert^2.
    \label{SPDCAS}
\end{equation}

\begin{figure}[b]
  \centering
  \includegraphics[width=0.48\textwidth]{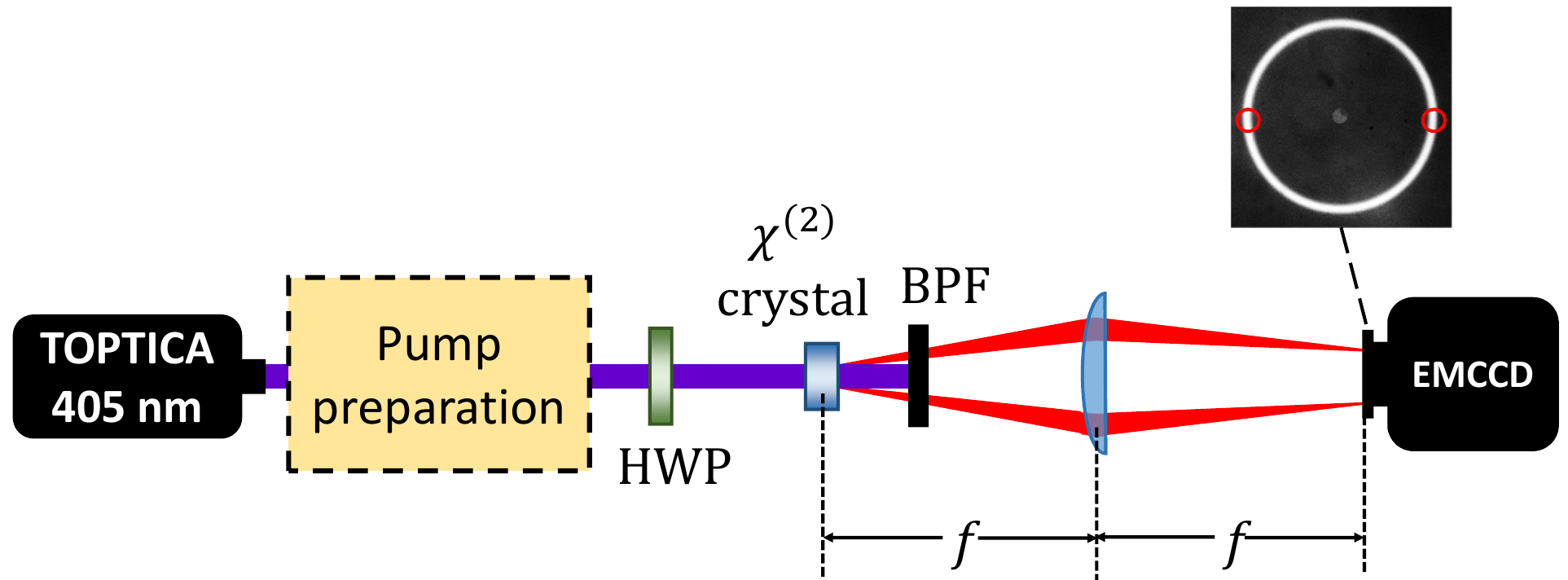}
  \caption{Experimental setup to record the angular spectrum of SPDC photons generated by pumping a non-linear crystal with different pump beams. Figure in inset is the SPDC annular distribution in the angular spectrum. Two diametrically opposite points (circled in red) on the distribution are chosen to collect the signal-idler pairs.}
  \label{exptalsetup_imaging}
\end{figure}

The experimental setup to record the angular spectra of SPDC photons with different helical pump modes is given in Fig. \ref{exptalsetup_imaging}. A pump beam of wavelength 405$\pm$2~nm from a continuous-wave diode laser (TOPTICA iBeam Smart) of 50~mW power is incident on a $\chi^{(2)}$ crystal. The dashed box corresponds to the case-by-case method to prepare NOV or POV helical beams using the method given in Fig. \ref{pump_prep_POV}(a). A half-wave plate (HWP) is used to orient the pump polarization along the optic axis of the crystal. As discussed earlier, we use Type-I BBO crystal to generate SPDC photon-pairs. The down-converted photons, signal \& idler, are generated in a non-collinear fashion at diametrically opposite points of the SPDC ring. A bandpass filter (BPF) of width 10~nm centered at 810~nm is used to filter down-converted photons and block the unconverted pump beam after the crystal. A $2f$-imaging configuration with a plano-convex lens of focal length 50~mm is used to image the SPDC in $k$-space. The angular spectrum of SPDC is recorded using an electron-multiplying CCD (EMCCD) camera with a gain $\times$100 and the addition of 100 frames, each having an exposure time of 0.5~s. The EMCCD camera has an imaging area of 512$\times$512 pixels with a pixel size of 16~$\mu$m.

\begin{figure}[h]
	\centering
    \includegraphics[width=0.48\textwidth]{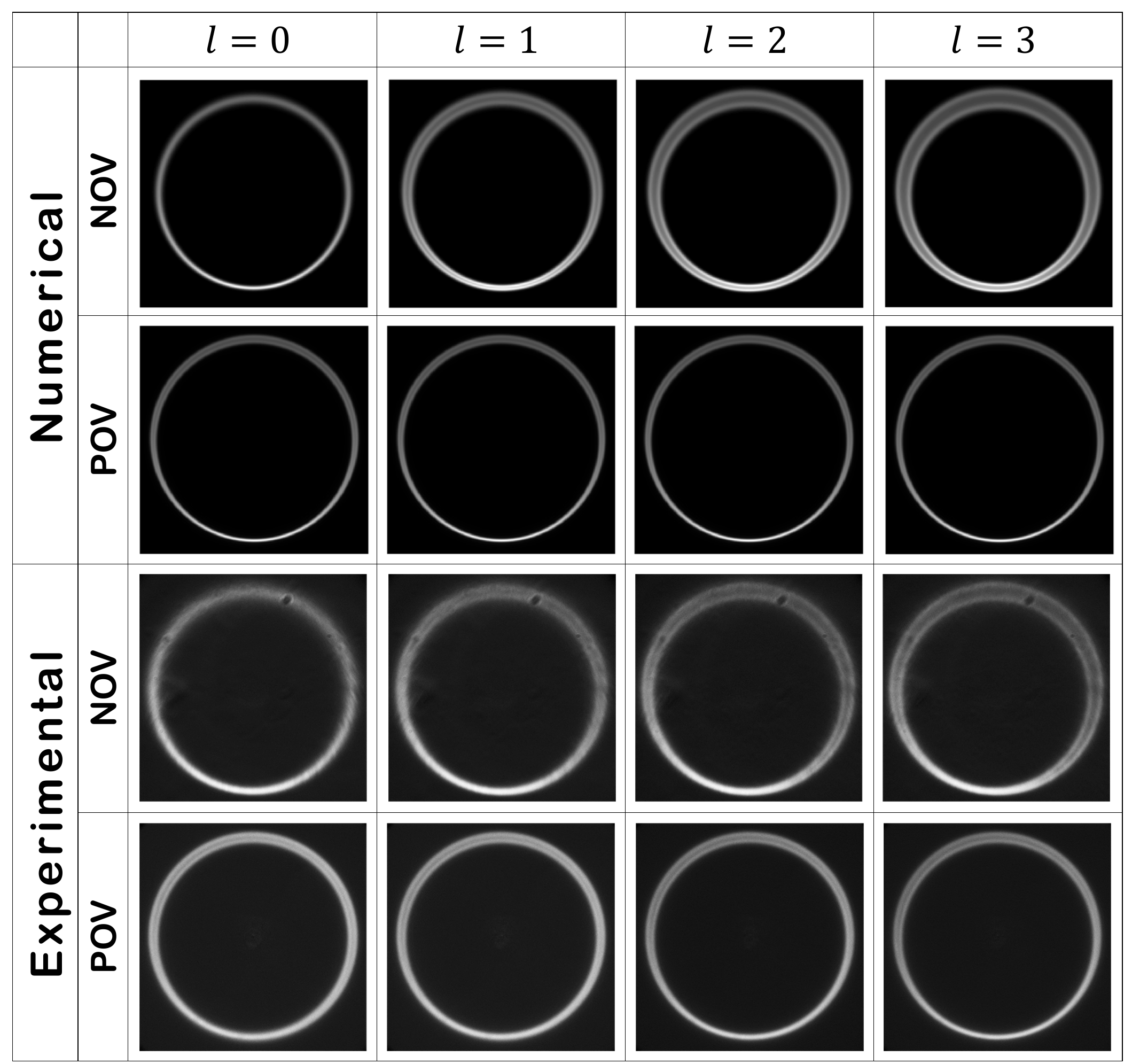}
	\caption{Numerical (top) and experimental (bottom) angular spectra of SPDC generated using NOV, and POV pump modes of various orders.} \label{SPDC-angular-spectra}
\end{figure}

The numerical angular spectra of SPDC with different pump OAM modes with similar experimental conditions obtained from Eqn. \eqref{SPDCAS} is shown in Fig. \ref{SPDC-angular-spectra}(top). This is well-matched with their corresponding recorded intensity distributions of SPDC photons with NOV and POV pumps of different OAMs, as shown in Fig. \ref{SPDC-angular-spectra}(bottom). The asymmetric intensity distribution for all the angular spectra is mainly due to the spatial walk-off of the thick crystal used. For a NOV pump, the SPDC annular distribution broadens with an increase in the topological order. However, the signature of 'size-invariance' of POV pump beams is observed in their corresponding angular spectra, where the intensity distribution remains the same for higher OAM values. 

For a non-collinear down-conversion, the signal and idler are located at diametrically opposite points of the annular distribution (Fig. \ref{exptalsetup_imaging}). If one chooses to place a collection optics setup each at such two points and detect the photons, then we could observe a decrease in the number of detected photons with the increase in OAM of an NOV pump. There will not be a significant change in the number of detected photons in the case of a POV pump. To achieve the full advantage of our study, we preferred the pump generation method illustrated in Fig.\ref{pump_prep_POV}(a) over Fig.\ref{pump_prep_POV}(b) due to the limitation of spiral-phase plates of order more than 3. As the low down-conversion rates in BBO crystal restricted us to investigate further the size effects on photon-pair detection, we choose periodically poled KTP (ppKTP) crystal, which not only provides higher photon-pairs production rate, however also removes the asymmetry in the SPDC distribution.
Pumping the crystal with a beam carrying OAM $l_p$ and the collection of idler photon through a single-mode fiber leads to a conditional detection of signal photon (heralded) carrying an OAM identical to that of the pump, which is discussed below.

\subsection{Heralded twisted single photons}

Equation \eqref{modefna} gives the two-photon modal distribution for a pump with electric field $E_p(\mathbf{k}_p^{\perp})$. According to conservation of OAM in SPDC process \cite{walborn2004entanglement}, for a pump carrying an OAM $l_p$ and the idler photon projected to a zero-OAM mode ($l_i=0$), the corresponding heralded signal photon will have the same OAM as that of the pump ($l_s=l_p$). This heralded single-photon carrying OAM can be experimentally generated by coupling one photon in a pair to a single-mode fiber and the other one to a multimode fiber. So, the coincidence counts can be calculated from the modal distribution as
\begin{equation}
    C \propto \int_0^{a_s}\int_0^{a_i}\Phi(\mathbf{k}_s^{\perp},\mathbf{k}_i^{\perp},\Delta k)\xi_s^*(\mathbf{k}_s^{\perp})\xi_i^*(\mathbf{k}_i^{\perp})d\mathbf{k}_s^{\perp}\mathbf{k}_i^{\perp},
    \label{coinc_counts_herladed}
\end{equation}
where $\xi_s$ and $\xi_i$ are the characteristic functions of spatial mode projectors in signal and idler arms respectively. The fibers in each arm is represented by a Gaussian function with mode-field diameters $2a_s$ and $2a_i$, given by
\begin{align}
    \xi_{s,i}^*(\mathbf{k}_{s,i}^{\perp})=\sqrt{\frac{a_{s,i}^2}{2\pi}}\text{exp}\left(-\frac{a_{s,i}^2}{4}|\mathbf{k}_{s,i}^{\perp}|^2\right).
    \label{FiberExpression}
\end{align}

\begin{figure}[h]
  \centering
  \includegraphics[width=0.48\textwidth]{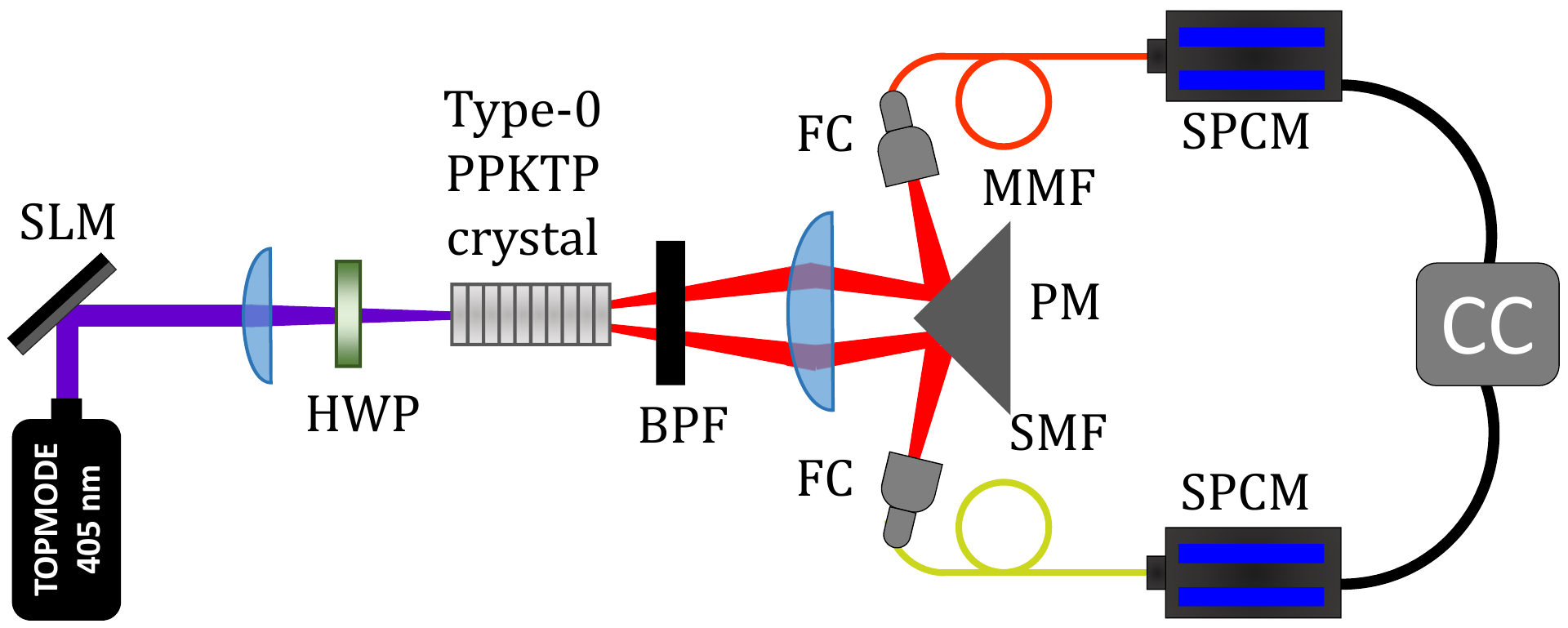}
  \caption{Experimental setup for generating heralded twisted single photons. BPF - Band pass filter; PM - Prism mirror; SMF - Single mode fiber; MMF - Multimode fiber; FC - Fiber coupler; SPCM - Single photon counting module; CC - Coincidence counter.}\label{exptalsetup_her_twist}
\end{figure}

The experimental setup to generate heralded twisted single-photons from SPDC is given in Fig. \ref{exptalsetup_her_twist}. Here, we have used a blue diode laser (TopMode) of wavelength 405~nm and power 10~mW with a spectral bandwidth of 0.1~nm, as the pump beam. NOV and POV beams are generated by illuminating the Gaussian laser mode onto the corresponding grating holograms imprinted on a SLM (Hamamatsu), and the first order far-field diffracted beam after Fourier transformation using a 750~mm lens is incident on a Type-0 ppKTP crystal of thickness 30mm and transverse dimensions of 1~mm$\times$2~mm. The HWP allows us to vary the pump beam polarization along the crystal axis. 

The down-converted photons (signal \& idler) of wavelength 810 nm each (degenerate) are generated in a non-collinear fashion at diametrically opposite points of the SPDC ring. To measure the number of generated photon pairs, two diametrically opposite portions of the SPDC ring at a given plane were selected using apertures (not shown in the setup) and the photons coming out of each aperture were collected using fiber collimators (CFC-2X-B, Thorlabs), of focal lengths 2 mm each. The fiber collimator in the idler arm is attached to a SMF (P1-780A-FC-2, Thorlabs) having a numerical aperture of 0.13 and a mode field diameter of $5\pm0.5$ $\mu$m, and that in the signal arm is attached to a multi-mode fiber (M43L02, Thorlabs). The fibers are connected to the single-photon detectors SPCMs (SPCM-AQRH-16-FC, Excelitas). The detectors have a timing resolution of 350 ps with 25 dark counts per second. To count the correlated photon-pairs, the two detectors are connected to a coincidence counter (IDQuantique-ID800), having a time resolution of 81 ps.

\begin{figure}[h]
  \centering
  \includegraphics[width=0.48\textwidth]{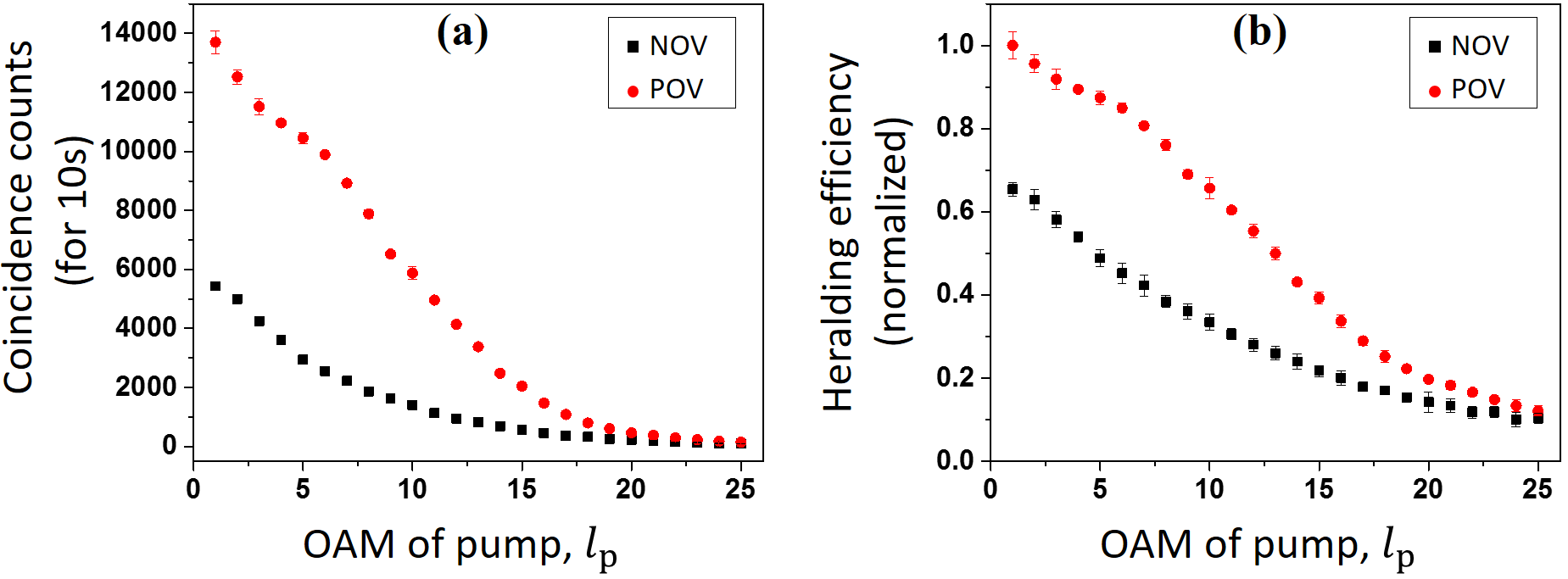}
  \caption{Plot of (a) measured coincidence counts for different values of OAM of pump, and (b) corresponding heralding efficiency, for NOV and POV pumps.}\label{plot1}
\end{figure}

We recorded the coincidences of signal and idler for NOV and POV pumps starting from $\ell=1$ to 25. Figure \ref{plot1}(a) shows the coincidence counts for NOV and POV pump beams with different OAM. For a pump OAM up to 10, the coincidence counts of SPDC photon pairs is considerably higher for a POV pump than that with a NOV pump beam. 

The difference in coincidences corresponding to the two pump modes of the same order decreases for higher orders. The robustness of the source with POV pump mode can be more understood by comparing the heralding efficiencies. Figure \ref{plot1}(b) shows the variation of heralding efficiency of SPDC single photons with pump OAM for NOV and POV pumps. From the graph, it is clear that, for up to a pump OAM of 15, the efficiency of heralded single photons carrying OAM with POV pump is considerably greater than that corresponding to the NOV pump. Ideally, the coincidences for POV pump should be independent of the order. However, we observe a drop in their coincidences for higher OAM values. This is due to the fact that the ring diameter of the POV mode gradually increases with order and the difference is more exposed for orders starting from $l=15$, which causes a considerable drop in counts.

\section{Two-photon OAM states using POV pump} \label{high-dim-OAM-states}
\subsection{Theory}

Parametric down-converted photons generated using a pump beam of any spatial profile will have an incoherent sum of different OAMs. Based on the OAM conservation in SPDC process \cite{walborn2004entanglement}, the photon-pairs represent a joint signal-idler OAM state in the allowed OAM subspace, given as
\begin{equation}
    \ket{\psi}=\sum_{\ell_i=-\infty}^\infty C_{\ell_p-\ell_i,\ell_i}\ket{\ell_p-\ell_i}_a\ket{\ell_i}_b,
    \label{SPDC_single_sum}
\end{equation}
where $l_i$ is the OAM of idler photon, $C_{l_p-l_i,l_i}$ is the probability amplitude for occurrence of the state $\ket{l_p-l_i}_a\ket{l_i}_b$. The subscripts $a$ \& $b$ represents signal and idler modes respectively. In the experiment, the probability amplitude corresponds to the coincidence counts obtained by projecting conjugate azimuthal modes in signal and idler. The coincidence counts representing each projection is given by
\begin{equation}
    C_{\ell_s,\ell_i}\propto {}^{}_{a}\langle \ell_s^{\text{(proj)}}\vert {}^{}_{b}\langle \ell_i^{\text{(proj)}}\vert \psi\rangle.
    \label{prob_ampl}
\end{equation}
Here, $\ell_s^{\text{(proj)}}$ and $\ell_i^{\text{(proj)}}$ are the OAM values of azimuthal modes considered in projective measurement. 

For a nearly collinear phase-matched SPDC, the probability amplitude in Eqn. (\ref{prob_ampl}) is rewritten in terms of an integral in polar coordinates representing the field overlap of pump, projected signal and idler modes
\begin{equation}
    C_{\ell_s,\ell_i}\propto\int_0^{2\pi}d\theta\int_0^\infty rE_p(r,\theta)E_s^*(r,\theta)E_i^*(r,\theta)dr,
    \label{prob_ampl_num}
\end{equation}
where $E_p$ is the pump mode and $E_s$ \& $E_i$ are the projected signal and idler modes respectively. Here, we consider a POV pump represented by \eqref{POV-expression} and NOV or BGV modes in both signal and idler arms. 

\subsection{Experiment}
\begin{figure}[h]
  \centering
  \includegraphics[width=0.48\textwidth]{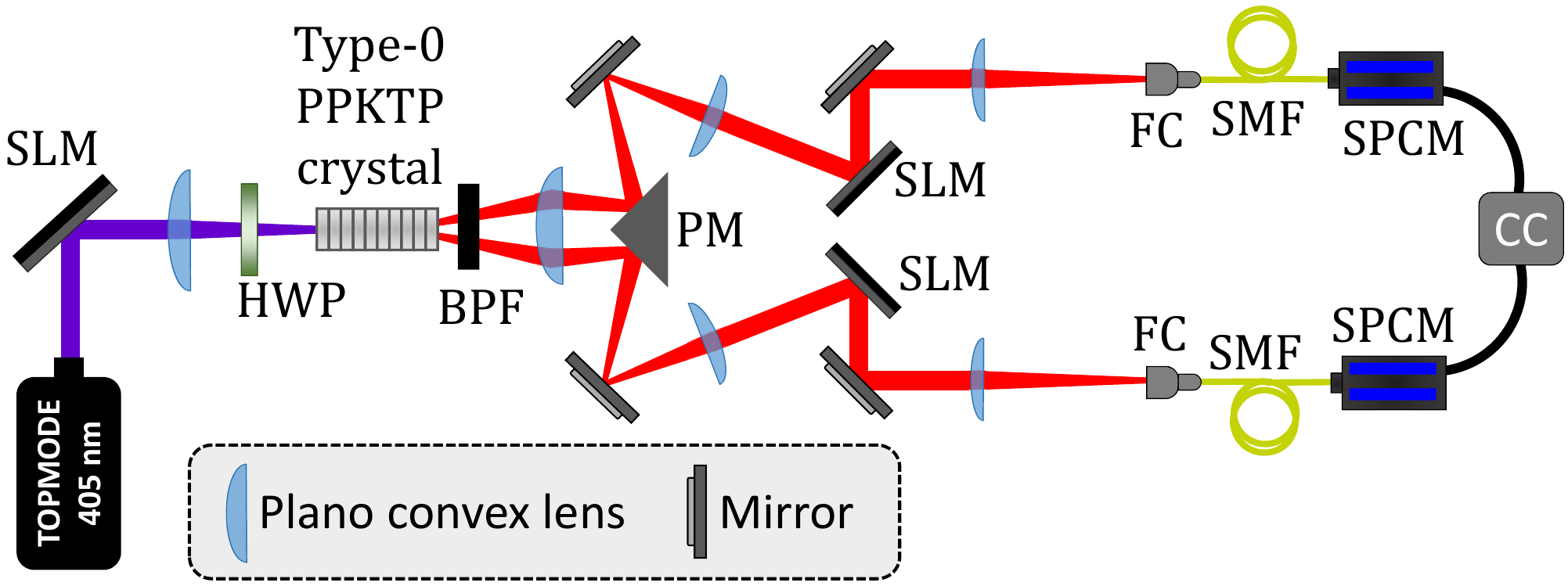}
  \caption{Experimental setup for generating OAM entangled photon-pairs with different helical pump modes.}\label{OAMentlsetup}
\end{figure}

The schematic of the experimental setup to measure in higher dimension is shown in Fig. \ref{OAMentlsetup}. In this case, the combination of a SLM and a SMF was used as the detection system, instead of the heralding configuration using fibres as discussed earlier. The combination was added in both the signal \& idler arms. The SLM is used to generate either the NOV or POV modes at the plane of the crystal. The crystal is then imaged on the two SLMs using two lenses and then imaged again to the SMF. The photon is then sent to the APDs, and the signal is analysed using the coincidence logic (CC). 

\begin{figure}[h]
  \centering
  \includegraphics[width=0.48\textwidth]{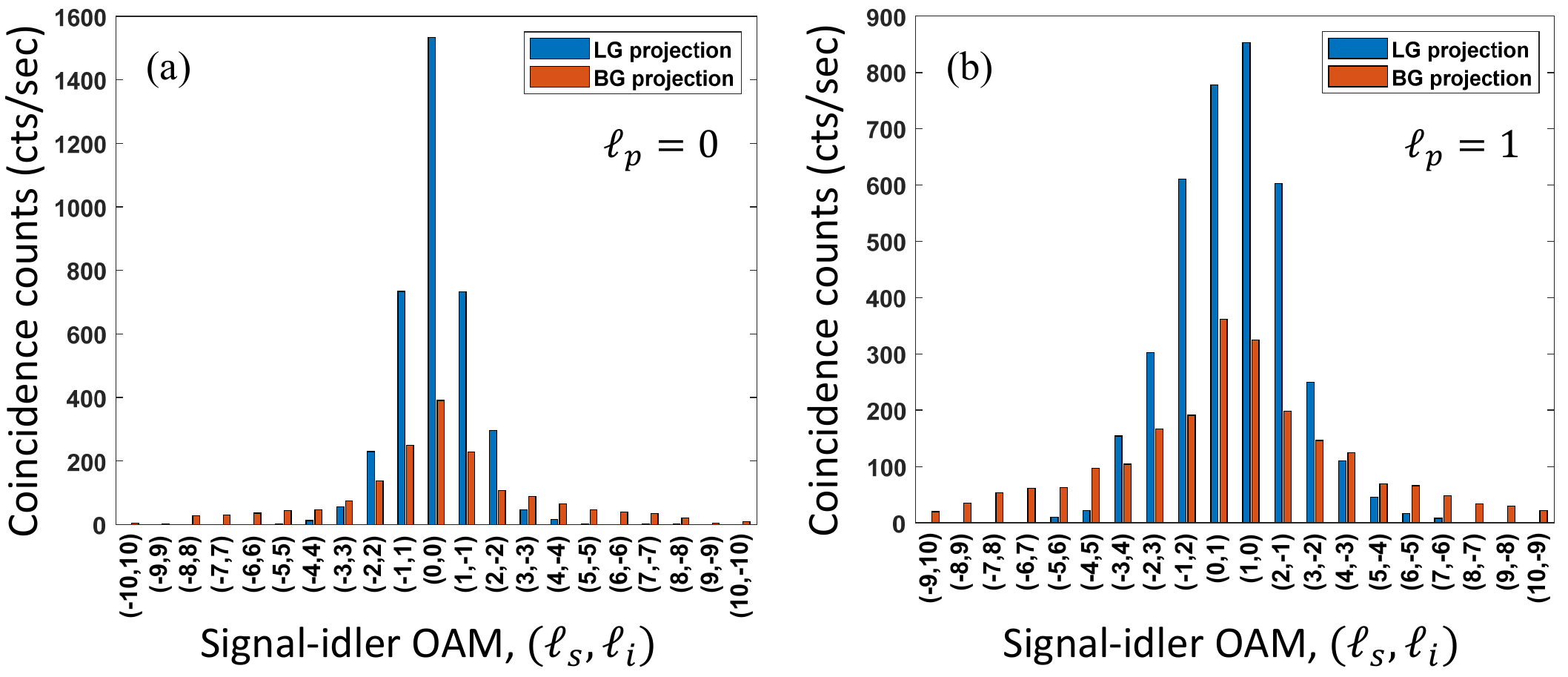}
  \caption{OAM spectrum of SPDC photon-pairs generated using POV pump of OAM (a) $\ell_p=0$ and (b) $\ell_p=1$.}\label{OAMspectra}
\end{figure}

We measured OAM spectra of SPDC photon-pairs with a POV pump and estimated their bandwidths using LG and BG mode projections in signal/idler. The OAM bandwidth is quantified as Schmidt number, which estimates the dimensionality of the state generated \cite{pors2008shannon, pires2010, straupe}. Fig. \ref{OAMspectra} shows the OAM spectra of photon-pairs for POV pump of order 0 \& 1. 
For $\ell_p=0$ (Fig. \ref{OAMspectra}(a)), the Schmidt numbers are 3.8 \& 8.7 for LG \& BG mode projections respectively. For $\ell_p=1$ (Fig. \ref{OAMspectra}(b)), the Schmidt numbers are 6.3 \& 11.6 for respective LG and BG mode projections. This shows that projecting the photon-pairs generated from a POV pump to a BG mode will effectively bring down the 'size effects' that causes a reduction in OAM bandwidth. Studies had already shown that the bandwidth could be further increased by increasing the radial wavenumber $k_r$, which broadens the spectrum further. \cite{mclaren2012entangled}.
\begin{figure}[h]
  \centering
  \includegraphics[width=0.48\textwidth]{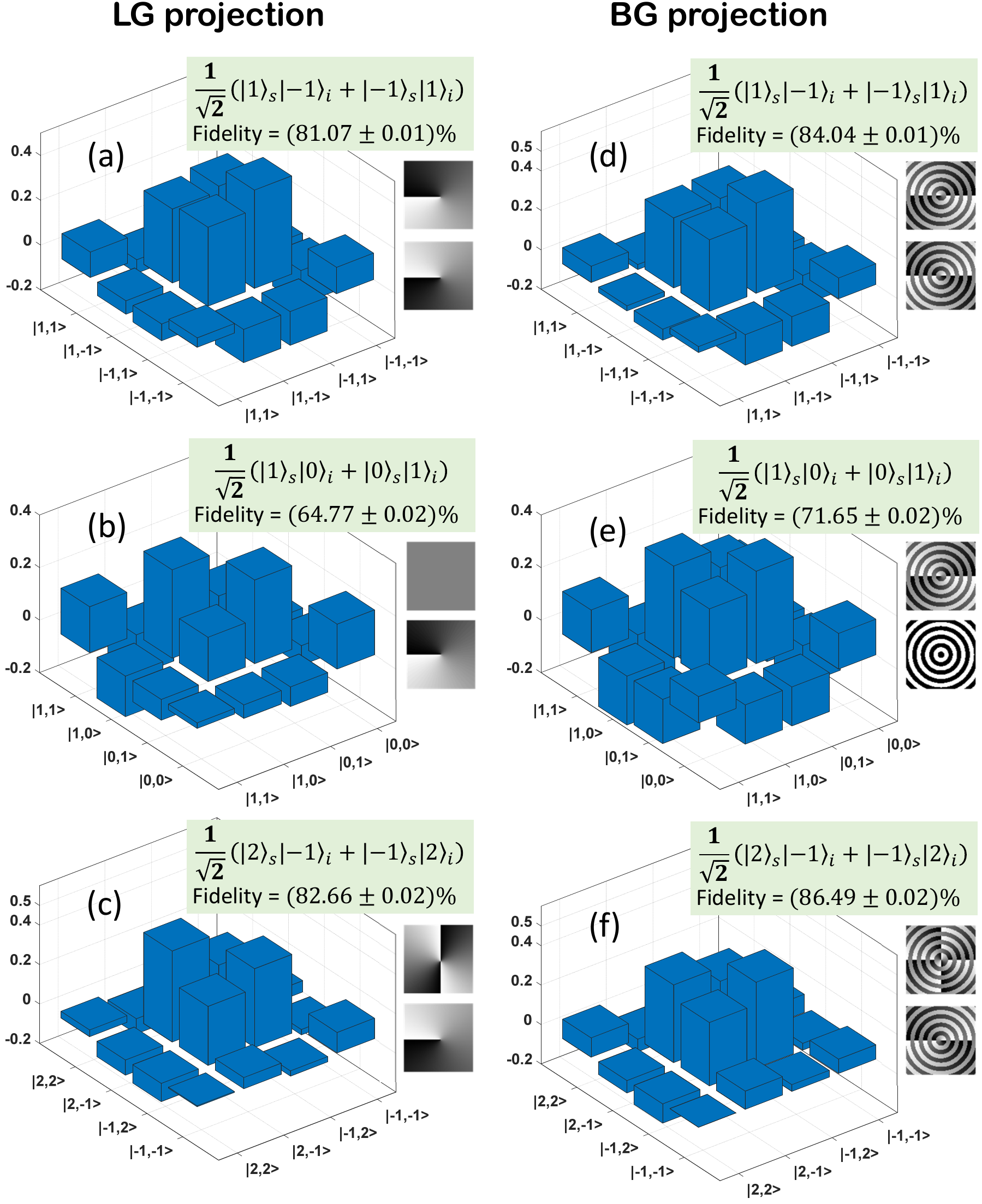}
  \caption{The density matrices of the two-photon OAM entangled states. (a)-(c) correspond to the estimation by projecting photon-pairs on LG basis, and (d)-(f) are the estimations in BG projection. The holograms or the true state and the fidelity of the estimated state are given in the inset of each density matrix. (a) \& (d) corresponds to a POV pump of $\ell_p=0$ and the other density matrices correspond to a POV pump of $\ell_p=1$. }\label{densitymatrix-2doamstates}
\end{figure}

We also studied the effect of zero-order mode projection in idler on the fidelity of generated quantum states using POV pump. For simplicity, we consider two-dimensional OAM states in the form of Bell states $(\ket{\ell_s}\ket{\ell_i}+\ket{\ell_i}\ket{\ell_s})/\sqrt{2}$, with different $\ell_s$ and $\ell_i$ values. Figure \ref{densitymatrix-2doamstates}(a)-(c) show the density matrices of OAM states estimated with LG mode projection and Fig. \ref{densitymatrix-2doamstates}(d)-(f) gives the corresponding density matrices with BG mode projections. In all these cases, the pump OAM $\ell_p$ is the sum of OAMs of signal and idler, $\ell_p=\ell_s+\ell_i$. Here, we observe that the state fidelity is lower for the case of a zero-OAM mode in signal or idler, which is the setting for the generation of heralded twisted single photons, as discussed in the earlier section. So, there is a trade-off in using a two-photon SPDC system with complete mode projection capabilities and using heralding with the fiber projective systems. 

We also observed that the fidelity of the state is improved with the use of BG projection. Efficient OAM detection based on phase-flattening of POV modes in classical light source has been demonstrated \cite{pinnell2019quantitative, pinnell2019perfect}. Thus, a robust source of high dimensional quantum states in OAM degree of freedom can be implemented by using a size-invariant twisted optical mode like a perfect optical vortex as a pump, as well as by performing projection of photon-pairs in Bessel-Gaussian modes.

\section{Conclusion} \label{Conclusion}
In conclusion, we have experimentally demonstrated that the conditional coupling efficiency of the heralded twisted single photons for higher OAM values can be improved by using a perfect optical vortex beam as the pump. We also showed that the dimensionality of the two-photon OAM states is increased with the use of POV modes in the pump, as well as projective measurements using Bessel-Gaussian vortex modes that give POV, instead of the Laguerre-Gaussian modes. The presented results may be utilized for the practical realization of efficient higher dimensional OAM entangled photon-pair sources. 

\section*{Disclosures}
The authors declare that there are no conflicts of interest related to this article.

\bibliography{Reference}

\end{document}